\documentclass[twocolumn]{aastex631}

\newcommand\Gt{G292.0$+$1.8}
\newcommand\M{$M_{\rm{ZAMS}}$}
\shorttitle{AASTeX v6.3.1 Sample article}
\shortauthors{Narita et al.}
\graphicspath{{./}{figures/}}

\begin{document}

\title{Evidence for Type Ib/c origin of the supernova remnant \Gt}

\author{Takuto Narita}
\affiliation{Department of Physics, Kyoto University Kitashirakawa Oiwake-cho, Sakyo, Kyoto, 606-8502, Japan}

\author{Hiroyuki Uchida}
\affiliation{Department of Physics, Kyoto University Kitashirakawa Oiwake-cho, Sakyo, Kyoto, 606-8502, Japan}

\author{Jacco Vink}
\affiliation{Anton Pannekoek Institute \& GRAPPA, University of Amsterdam, Netherlands}

\author{Satoru Katsuda}
\affiliation{Graduate School of Science and Engineering, Saitama University, Sakura, Saitama, Japan}

\author{Hideyuki Umeda}
\affiliation{Department of Astronomy, Graduate School of Science, the University of Tokyo, Tokyo, 113-0033, Japan}

\author{Takashi Yoshida}
\affiliation{Yukawa Institute for Theoretical Physics, Kyoto University, Kitashirakawa Oiwake-cho, Sakyo, Kyoto 606-8502, Japan}

\author{Toshiki Sato}
\affiliation{Department of Physics, School of Science and Technology, Meiji University, 1-1-1 Higashi Mita, Tama-ku, Kawasaki, Kanagawa 214-8571, Japan}

\author{Kai Matsunaga}
\affiliation{Department of Physics, Kyoto University Kitashirakawa Oiwake-cho, Sakyo, Kyoto, 606-8502, Japan}

\author{Takeshi Go Tsuru}
\affiliation{Department of Physics, Kyoto University Kitashirakawa Oiwake-cho, Sakyo, Kyoto, 606-8502, Japan}

\begin{abstract}
Circumstellar material (CSM) produced by mass loss from massive stars ($\gtrsim10~M_{\odot}$) through strong stellar winds or binary stripping provides rich information for understanding progenitors of core-collapse supernova remnants.
In this paper we present a grating spectroscopy of a Galactic  SNR \Gt, which is claimed to be a Type Ib/c remnant in a binary system according to recent studies.
If \Gt \ was experienced a strong mass-loss via binary interactions before its explosion, an oxygen-rich material produced in the He-burning layer is expected to be observed in the central belt-like structure formed by shock-heated CSM.
Using the Reflection Grating Spectrometer onboard XMM-Newton, we detect \ion{N}{7} Ly$\alpha$ line (0.50~keV) for the first time in \Gt \ and find that the abundance ratio of nitrogen to oxygen is significantly lower (N/O$=0.5\pm0.1$) than the solar value.
This low N/O suggests that the progenitor of \Gt \ experienced strong mass-loss and ended up to a Wolf-Rayet (WR) star exposing the He-burning layer at the pre-supernova.
Comparing our result and the evolution models of single stars and binaries, we conclude that the progenitor of \Gt \ experienced strong mass-loss enough to occur a Type Ib/c supernova.
Our finding is another crucial piece of evidence for a stripped-envelope supernova such as Type Ib/c as the origin of \Gt.
\end{abstract}

\section{Introduction} \label{sec:intro}
Circumstellar material (CSM) around core-collapse (CC) supernovae (SNe) of massive ($\gtrsim10~M_{\odot}$) stars have rich information on  progenitor parameters such as a mass of zero age main sequence (\M), an initial rotation velocity, and convection \citep[e.g,][]{Fransson_2005, Maeder_2014, Chiba_2020}.
This is because the mass-loss history of massive stars is affected by their stellar properties; particularly abundances of the CSM is significantly changed during the stellar evolution \citep[e.g,][]{Maeder_1983}.
For instance, main-sequence (MS) massive stars emit a hydrogen(H)-rich wind and post-MS stars emit a stellar wind rich in helium (He) and nitrogen (N), since elements such as N produced by the CNO-cycle in the H-burning layer are carried to the surface by the convection \citep[e.g,][]{Owocki_2004, Przybilla_2010}.
\par

CC SNe are roughly classified by the presence of lines, i.e., the abundance of the CSM; H-rich (Type II), H-poor (Type Ib), and H/He-poor (Type Ic) SNe.
Among them, Type Ib/c SNe are also known as stripped SNe because the progenitors are Wolf-Rayet (WR) stars experiencing strong mass-loss and exposing inner layers where the H/He burning takes place \citep[e.g,][]{Smith_2014}.
Therefore, the progenitor of Type Ib/c SNe is expected to have blown the N-rich winds produced in the H-burning layer and oxygen (O) rich winds produced by the triple-$\alpha$ reaction and $^{12}$C($\alpha$, $\gamma$)$^{16}$O reaction in the He-burning layer.
Although the CSM around SNe  thus gives us rich information on their progenitors, due to a lack of observational constraints, many questions still remain about the evolution of massive stars and their explosion mechanisms.
 \par

Supernova remnants (SNRs) are also essential to determine the origin of CC SNe, since abundances of shock-heated ejecta and CSM reflect progenitor parameters such as stellar mass and mass loss, which are useful to constrain their explosion scenario, i.e., their progenitors or progenitor systems.
The most common method employed so far to estimate \M \  is to compare  ejecta abundance ratios of such as O, neon (Ne), magnesium (Mg), silicon (Si), sulfur (S), and iron (Fe) with theoretical calculations \citep[cf.][]{Katsuda_2018}.
\par

Lighter elements contained in the CSM, Carbon (C), N, and O (hereafter, CNO elements) are also useful to constrain the progenitors or progenitor systems.
Several previous observations with the Reflection Grating Spectrometer (RGS) onboard XMM-Newton successfully detected \ion{C}{5}~Ly$\alpha$ and \ion{N}{7}~Ly$\alpha$ in nearby SNRs \citep[e.g,][]{Uchida_2019, Kasuga_2021}.
Once we can accurately measure the amount of these CNO elements in the CSM, we can directly access the progenitor's mass-loss history since the mass-loss rate at each stage of the stellar evolution changes the amount of CNO elements blown out from the stellar surface \citep[e.g,][]{Maeder_1983}, which will be a good tracer of progenitor parameters or progenitor systems \citep[e.g,][]{Chiba_2020, Tateishi_2021}.
Under this consideration, in our previous paper we established a new method for constraining these parameters using a high-energy-resolution spectroscopy and determined \M \ and the rotation velocity of a magnetar-hosting SNR RCW~103 \citep{Narita_2023}.
\par

Here, we focus on an oxygen-rich CC SNR \Gt, which has a belt-like structure produced by CSM \citep{Park_2004, Park_2002} and hosts a rapid moving pulsar {\citep{Hughes_2001, Long_2022}.
Previous studies show various estimates for \M \ of the progenitor with ejecta abundances: 25--40~$M_{\odot}$\citep[][]{Park_2004, Gonzalez_2003, Kamitsukasa_2014}, 13--30~$M_{\odot}$\citep[][]{Bhalerao_2019}, $<15~M_{\odot}$\citep[][]{Katsuda_2018}., and 12--16~$M_{\odot}$\citep[][]{Temim_2022}.
Interestingly a recent study of hydrodynamical simulations of \Gt \  suggests that a total mass of the ejecta requires $\lesssim3~M_{\odot}$, which likely results in a remnant of a massive star in a binary system \citep{Temim_2022}.
\par

In this paper, we perform a grating spectroscopy of  CSM-dominant filaments in \Gt \ (Section~\ref{sec:obs} and \ref{sec:ana_results}), estimate the progenitor in the cases of a single star or a binary system (Section~\ref{subsec:CSM_comp} and \ref{subsec:N/O}), and then summarize the conclusions in Section~\ref{sec:con}.
Throughout the paper, we assume the age and distance of \Gt \ to be 3000~yr \citep{Ghavamian_2005, Winkler_2009} and 6.2~kpc \citep{Gaensler_2003}, respectively.
Errors of parameters are defined as 1$\sigma$ confidence intervals.
\par

\section{observations} \label{sec:obs}
We selected an observation with the longest exposure time available for \Gt \ (ID: 0400330101) from the XMM-Newton Science Archive since it is inappropriate to combine the spectrum from different observations with different roll angles.
A nearby blank-sky observation (ID: 0823031601) was applied for estimating the background contribution.
RGS and MOS data were used for the following analysis.
We also used an ACIS image of \Gt \ obtained with Chandra (ID: 6677) for estimating the spatial broadening of the remnants to modify the RGS response.
We reprocessed the MOS data using pipeline tool \texttt{emchain} in the XMM-Newton Science Analysis System (SAS) version 20.0.0. 
The RGS data was reprocessed using the pipeline tool \texttt{rgsproc} in SAS. 
Remaining exposure times after the standard event screening are 60.7~ks and 61.0~ks for the MOS and RGS, respectively.

\begin{figure}[ht]
 \begin{center}
  \includegraphics[width=80mm]{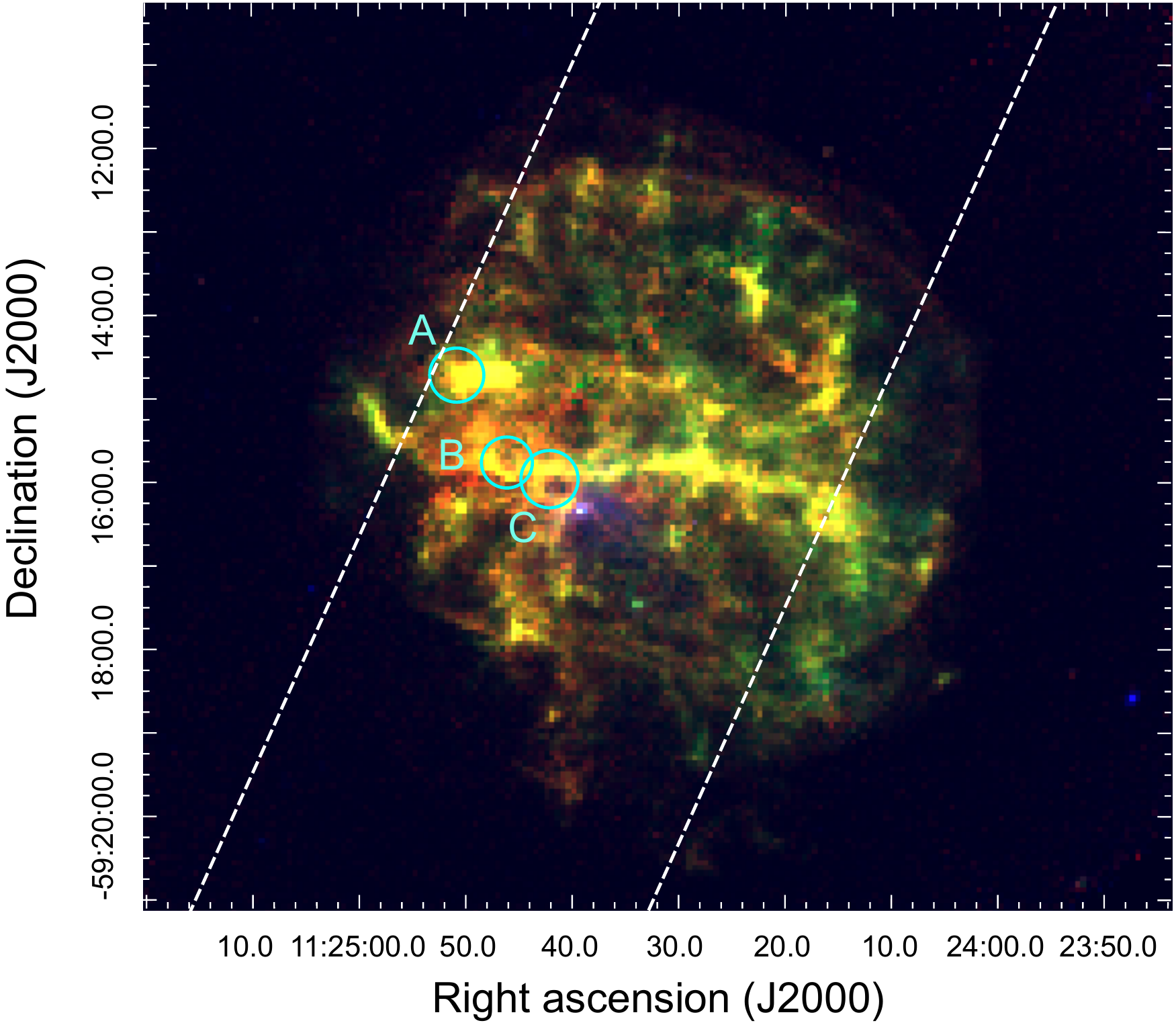}
 \end{center}
 \caption{True-color flux image of \Gt \  obtained with ACIS. Each color corresponds to each energy bands of 0.4--0.7~keV (red), 0.7--1.2~keV (green), 2.0--10.0~keV (blue). The white dashed lines represent the cross-dispersion area of the RGS and cyan circles indicate the spectral extraction regions.}
 \label{fig:rgb_image}
\end{figure}

\section{analyses and results} \label{sec:ana_results}
As shown in Figure~\ref{fig:rgb_image}, we observe a clear belt-like structure, which is thought to be a product of stellar winds  due to its low metal abundance compared to that  in the metal-rich ejecta elsewhere in the remnant \citep{Park_2002}.
We selected three compact bright regions in the belt-like filament (namely, Reg~A, B, and C in Figure~\ref{fig:rgb_image}) so that we can extract high resolution spectra with the RGS.
The angular size of each region is small ($\lesssim 1'$) enough to resolve low energy lines of the CNO elements.
\par

For the following spectral analysis, we used the \texttt{XSPEC} software version 12.12.1 \citep{Arnaud_1996} and the maximum likelihood W-statistic \citep{Wachter_1979}.
MOS1, MOS2, and combined RGS data (RGS1$+$2; first and second order spectra) were simultaneously fitted.
RGS response matrices (RMFs) were modified for taking into account the spatial broadening using \texttt{rgsrmfsmooth} in \texttt{FTOOL}.
We convolved the RMFs to an ACIS image for each emission line and consequently obtained 30 different modified RMFs applied for different emission line bands.
\par

We fitted the spectra with an absorbed two-component non-equilibrium ionization (NEI) model with variable abundances, \texttt{tbabs*(vnei+vnei)} since according to previous observations \citep[e.g,][]{Gonzalez_2003, Vink_2004, Bhalerao_2019} the analyzed regions are explained by emissions from both the forward-shocked CSM and reverse-shocked ejecta.
We set the solar abundance to that of \cite{Wilms_2000} defied as $({n_x}/{n_H})_{\odot}$, where $n_X$ and $n_H$ represent the number density of each element.
The hydrogen column density ($N_{\rm{H}}$) was set to free and the electron temperature ($kT_{\rm e}$), the ionization timescale ($n_{\rm e}t$), and the normalization for  each NEI component  were allowed to vary.
We also varied abundances of O (=Ne), Mg, Si, S, and Fe (=Ni) for the hot component and those of N, O, Ne, and Fe (=Ni) for cold components.
Here, we define "abundance" as the so-called solar abundance ratio: $(n_{\rm{x}}/n_{\rm{H}})/(n_{\rm{x}}/n_{\rm{H}})_{\odot}$.
We also added a power-law component to the two-component NEI models for Reg C since this region is contaminated by emission from the pulsar wind nebula \citep[e.g,][]{Kamitsukasa_2014}.
\par

We found that the above model was not able to represent the continuum in the high energy band ($\gtrsim5$~keV).
According to  the previous Suzaku observation of \Gt, \citet{Kamitsukasa_2014} suggests that the ejecta  can be explained by a multi-temperature model ($\sim 1.1$~keV and $\sim2.7$~keV).
We therefore added another NEI component to the above model, allowing  the normalization, $kT_{\rm e}$ and $n_{\rm e}t$ to be free (hereafter, three-component NEI model).
The abundance of Fe of this NEI component was linked to that of the above hot component and the other elements were fixed to the solar value.
The goodness of fit is improved (W-statistics value/d.o.f.: 8,651/8,586 to 8,597/8,579) without any other significant change in the other parameters.
We thus applied the three-component NEI model for the following analysis.
\par

Figure~\ref{fig:spec}  shows  the fits of three-component NEI model to the spectra.
Table~\ref{tab:par} displays the best-fit parameters.
The obtained $N_{\rm{H}}$, $kT_{\rm{e}}$, and $n_{\rm{e}}t$  are consistent with the pervious studies \citep[][]{Park_2002, Gonzalez_2003, Vink_2004, Kamitsukasa_2014, Bhalerao_2019}.
The abundances of medium $kT_{\rm{e}}$ ($\sim0.7~\rm{keV})$ and low $kT_{\rm{e}}$  ($\sim0.2~\rm{keV})$ components are similar to the previous measurements of the ejecta \citep[][]{Park_2002, Gonzalez_2003, Kamitsukasa_2014, Bhalerao_2019} and the CSM \citep{Park_2002, Kamitsukasa_2014, Bhalerao_2019}, respectively.
We found no evidence of line broadening, reported in the previous study for \ion{Ne}{10} Ly$\alpha$ \citep{Vink_2004}, which might be due to the difference of the spectral extraction regions or statistics.
Notably we detected the \ion{N}{7} line ($=0.50$~keV) for the first time in \Gt.
Measured N-rich abundance ratio (N/Fe$\geq1.3$) in at least two regions (Reg B and C) suggests that the low-$kT_{\rm{e}}$ component is likely from the forward-shocked CSM containing N-rich winds blown out from the progenitor.
We note that if the plasma emission is produced by H-poor and He-rich stellar winds, the abundance ratios of the elements to H of low-$kT_{\rm{e}}$ component is estimated $\sim 4$ times larger than the fitted values and is larger than 1 solar.

\par

\begin{deluxetable*}{llccc}
\tablenum{1}
\tablecaption{Best-fit parameters of the spectrum}
\label{tab:par}
\tablewidth{0pt}
\tablehead{
\colhead{Component} & \colhead{Parametars (unit)} & \colhead{Reg~A} & \colhead{Reg~B} & \colhead{Reg~C}}
\startdata  
Absorption & $N_{\rm{H}}$ (10$^{22}$~cm$^{-2}$)& $0.51\pm{0.03}$ & $0.587^{+0.002}_{-0.001}$ & $0.60\pm{0.01}$   \\
Low-temperature NEI & $kT_{\rm{e}}$ (keV) &  $0.43^{+0.05}_{-0.02}$ & $0.2039^{+0.0008}_{-0.0001}$ & $0.206\pm{0.001}$  \\
(CSM)& N &  $0.2\pm{0.1}$ & $0.8\pm{0.1}$ & $0.4\pm{0.1}$  \\
& O & $0.6\pm{0.1}$ & $1.2\pm{0.1}$ & $1.1\pm{0.1}$  \\
& Ne & $0.71^{+0.10}_{-0.04}$ & $1.79^{+0.03}_{-0.08}$ & $1.7^{+0.1}_{-0.2}$   \\
& Fe(=Ni) & $0.16^{+0.03}_{-0.02}$ & $0.50^{+0.03}_{-0.05}$ & $<0.1$  \\
& $n_{\rm{e}}t~(\rm{cm^{-3}~s})$ & $5^{+2}_{-1}\times10^{10}$ & $\lesssim 10^{13}$ & $\lesssim 10^{13}$  \\
& normalization & $0.016^{+0.005}_{-0.004}$ & $0.08\pm{0.01}$ & $0.097^{+0.002}_{-0.013}$  \\
Middle-temperature NEI & $kT_{\rm{e}}$ (keV) & $0.80^{+0.03}_{-0.02}$ & $0.73\pm{0.01}$ & $0.63\pm{0.01}$  \\
(ejecta)& O(=Ne) & $26^{+11}_{-8}$ & $4.5^{+0.1}_{-0.2}$ & $3.3^{+0.1}_{-0.2}$  \\
& Mg & $12^{+5}_{-3}$ & $3.4\pm{0.2}$ & $2.3^{+0.1}_{-0.1}$  \\
& Si & $9^{+2}_{-3}$ & $2.0\pm{0.1}$ & $1.23^{+0.05}_{-0.06}$  \\
& S & $8^{+3}_{-2}$ & $1.1\pm{0.1}$ & $0.2\pm{0.1}$   \\
& Fe (=Ni) & $4^{+2}_{-1}$ & $0.77^{+0.01}_{-0.03}$ & $0.49^{+0.02}_{-0.03}$   \\
& $n_{\rm{e}}t~(10^{11}\ \rm{cm^{-3}~s})$ & $2.4\pm{0.3}$ & $1.85\pm{0.02}$ & $2.5^{+0.1}_{-0.2}$   \\
& normalization & $0.0012^{+0.0002}_{-0.0003}$ & $0.0104^{+0.0002}_{-0.0001}$ & $0.0194^{+0.0004}_{-0.0008}$   \\  
High-temperature NEI & $kT_{\rm{e}}$ (keV) & $2.4^{+0.5}_{-0.4}$ & $4.1^{+0.3}_{-0.2}$ & $5.6^{+0.3}_{-0.3}$   \\
(ejecta)& $n_{\rm{e}}t~( \rm{cm^{-3}~s})$ & $1.2^{+0.2}_{-0.3}\times10^{11}$ & $6.1^{+0.7}_{-0.3}\times10^{10}$ &  $3.7^{+0.3}_{-0.2}\times10^{10}$   \\
& normalization & $0.004^{+0.002}_{-0.001}$ & $0.00134^{+0.00003}_{-0.00005}$ & $0.0194^{+0.0008}_{-0.0003}$  \\  
Power law & Photon index & -- & -- & $\leq1.4$ \\
& normalization ($\rm{photons~keV^{-1}~cm^{-2}~s^{-1}}$) & -- & -- & $\leq0.00009$   \\
\hline
& W-statistic/d.o.f. & 7,255/8,579 & 10,178/8,574 & 10,178/8,574 \\
\hline
\enddata
\tablecomments{The unit of the abundances is represented in the form of solar abundance ratio defined as $({n_x}/{n_H})_{\odot}$, where $n_x,~n_H$ represent the number density of each element. The normalization of the NEI model is defined as ${10^{-14}\over{4\pi[D_A(1+z)]^2}}\int n_en_HdV$, where $D_A$ is the angular diameter distance to the source (cm), $dV$ is the volume element (cm$^3$), and $n_e$ and $n_H$ are the electron and H densities (cm$^{-3}$), respectively.}
\end{deluxetable*}

\begin{figure*}[htbp]
 \begin{center}
  \includegraphics[width=180mm]{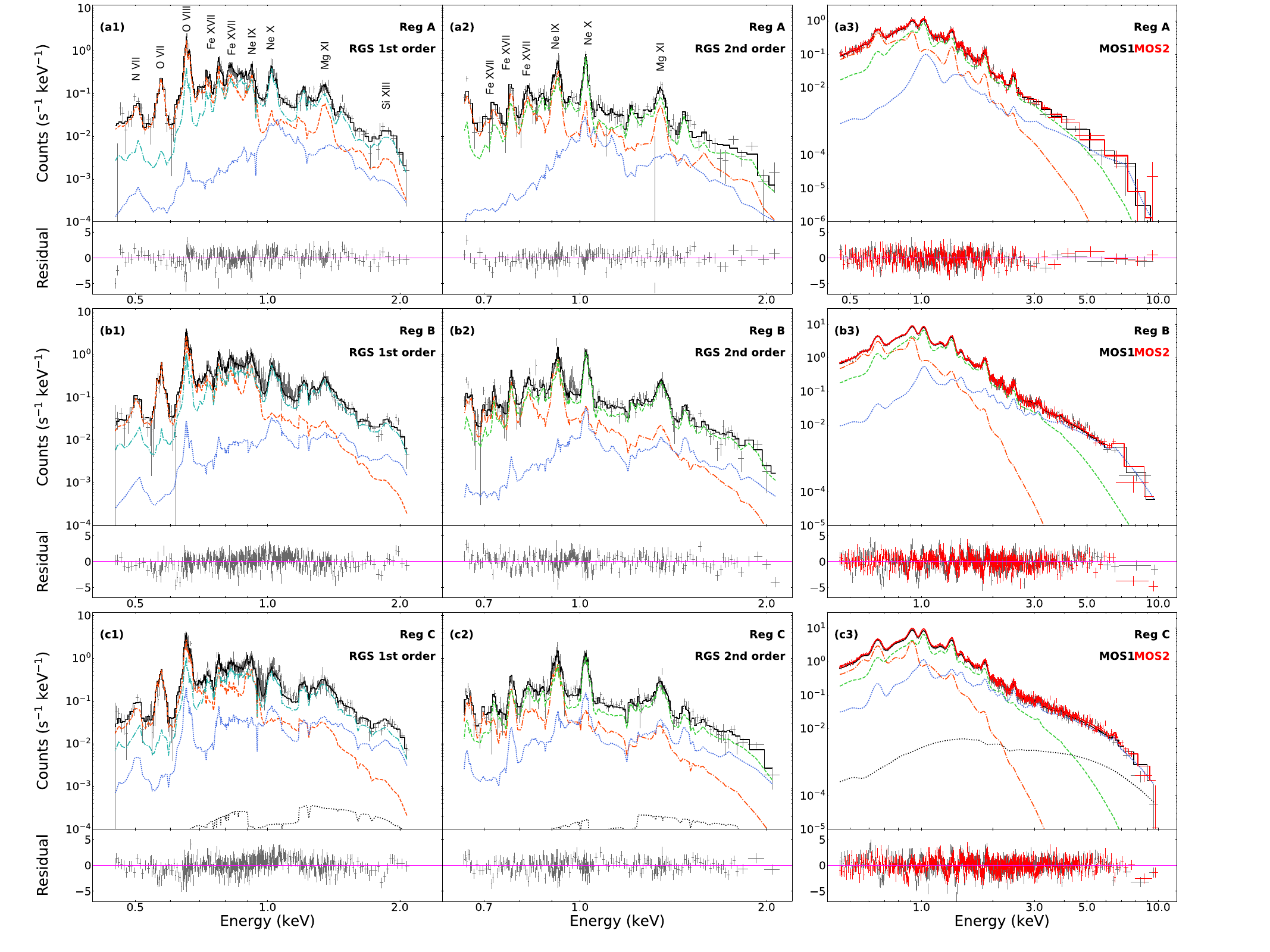}
 \end{center}
 \caption{X-ray Spectra of all the selected regions. \textit{Top}: Spectra of Reg~A obtained with the first- (a1) and second- (a2) order of RGS1$+$2 and those of MOS1/2 (a3). The orange dashed, green dot-dashed, and dotted lines are the best-fit NEI components with low, medium and high $kT_{\rm{e}}$, respectively. \textit{Middle and Bottom}: The same as the top, but spectra of Regs~B and C. The black dotted lines indicate the power-law component.}
 \label{fig:spec}
\end{figure*}

\section{discussion} \label{sec:dis}
\subsection{Evidence of CSM and its Composition}\label{subsec:CSM_comp}
Previous studies \citep[e.g,][]{Katsuda_2015, Tateishi_2021, Narita_2023}  claim that  the N-rich material observed in SNRs is clear evidence of CSM.
This is because N is mainly produced by the CNO cycle in the H-burning shell and blown out from the stellar surface via mass loss through the life of progenitors \citep[e.g,][]{Maeder_1983}.
Our results are consistent with the conclusion that the  N-rich material found in the belt-like structure of \Gt \ originates from the shock-heated CSM  as suggested by \citet{Park_2002}.
Such N-rich CSM  is presumably from a red super giant (RSG) or a WR star \citep[e.g,][]{Maeder_1983}.
\par

Recently we showed a method for constraining progenitor parameters such as \M, rotation velocities, metallicities, and convections by using the abundance ratio N/O of the CSM.
Initial state of progenitors changes the mass-loss rate in each stage of the stellar evolution having a positive relationship to the amount of C, N, and O of CSM.
In the case of relatively low \M ($\lesssim30~M_{\odot}$) progenitors, the N/O is higher than the solar value since the N-rich layer produced by CNO-cycle in H-burning layer is blown out from the stellar surface.
On the other hand, in the case of relatively high  \M ($\gtrsim30~M_{\odot}$) progenitors, the N/O is lower than the solar value since the O-rich layer produced by triple $\alpha$ in the He-burning layer is blown out from the stellar surface by the strong mass loss.
The N/O of rapid rotating progenitors is similar to that of heavier progenitors with slow initial rotation velocity since rotation enhances the mass-loss rate \citep[e.g.,][]{Heger_2000}.
Metallicity and convection also change the N/O since the metallicity has a positive relationship to the mass-loss rate \citep[e.g.,][]{Maeder_2014} and convection can carry the N-rich material to the outer layer of the progenitor \citep[e.g.,][]{Luo_2022}.
Under this consideration, we calculated the N/O by integrating the amounts of elements of stellar winds blown out from various type of progenitors with the stellar evolution codes and estimated the progenitor of RCW~103 (${\rm N/O}=3.8\pm0.1$) to be a low-mass ($M_{\rm{ZAMS}}=10$--12$M_\odot$) progenitor with medium rotation velocity ($\lesssim 100~\rm{km~s^{-1}}$) \citep{Narita_2023}.
In the same manner, we constrain the properties of the progenitor of \Gt \ for the following analysis. 
The weighted average N/O measured in the three regions is calculated to $0.5\pm0.1$ with confidence contour map between N and O of the soft component.
Of particular note is a significantly low N/O of \Gt \ ($=0.5\pm0.1$) compared with RCW~103, which implies that the progenitor  exposed the O-rich core before the SN explosion.
\par

For constraining the progenitor of \Gt, it is important to know what material is contained in the belt-like region.
The abundances of CSM produced by stellar winds are same to that produced by other mechanism such as the binary stripping since the abundances of the material producing CSM is uniform across the entire stellar surface.
Under this consideration, we constrained the CSM composition of \Gt \ assuming the spherically symmetric CSM.
Figure~\ref{fig:wind} illustrates two conceivable cases of  environments around the progenitor at the pre-supernova stage for \Gt.
In accordance with  \citet{Narita_2023}, we made a simple assumption  for calculating the bubble sizes by applying CSM hydrodynamic evolution \citep[e.g,][]{Chevalier_1999, Chevalier_2005} and  stellar evolution \citep[e.g,][]{Takahashi_2013, Takahashi_2014, Takahashi_2018, Yoshida_2019} codes of single stars.
In both cases, the outer most cavity is produced by MS winds, which sweep up the interstellar matter (ISM) in the MS phase \citep{Weaver_1977, Chen_2013}.
Assuming the small progenitor (\M$=10~M_{\odot}$) and the pressure divided by Boltzmann's constant of Galactic ISM \citep[$\sim10^4~\rm{cm^3~K}$;][]{Bergh_1998}, we calculated the minimum bound of the size of the wind bubble to be $\sim15.6~\rm{pc}$ since the bubble size has a positive linear relationship to \M \citep{Chen_2013}.
We also calculated the bubble size around a binary system with the hydrodynamic simulation of wind morphology around twin-star binaries \citep{MacLeod_2020} and binary evolution models \citep{Yoon_2010}.
From the mass-loss rate ($\sim10^{-8}~M_{\odot}~\rm{yr^{-1}}$) and the speed of wind ($\sim$300--1000~km~s$^{-1}$) emitted from the binary models, the stellar wind bubble size around a binary system is similar to that around a single star.
Since the radius of \Gt \ is $\sim8.0$~pc, we infer that the observed CSM component contain no ISM.
\par

We consider two cases with different CSM structures for estimating the swept-up stellar wind material in the observed CSM filaments.
In the case of a relatively low mass ($\lesssim20~M_{\odot}$) progenitor, the RSG wind material is concentrated near the progenitor ($r_{\rm{RSG}}=2.5$~pc) with a radius of $r_{\rm{RSG}}=2.5$~pc due to a slow wind \citep[$\sim10~\rm{km~s^{-1}}$,][]{Goldman_2017}.
We thus consider that the CSM component originates from a swept-up material produced by mass loss through the RSG phase.
In the case of a relatively high mass ($\gtrsim 20~M_{\odot}$) progenitor, a fast material produced by strong mass loss through the WR phase \citep[100--200~$\rm{km~s^{-1}}$,][]{Chevalier_1983, Chevalier_2005} expands up to $r_{\rm{WR}}=8.0$~pc ($\geq 60~M_{\odot}$) and sweeps up the RSG mass-loss materials.
Given this, we assume that the resultant CSM component contains either all of the RSG/WR mass-loss material produced or only a portion of the WR mass-loss material if the progenitor was a high-mass star. 
\par

\begin{figure*}[htbp]
 \begin{center}
  \includegraphics[width=180mm]{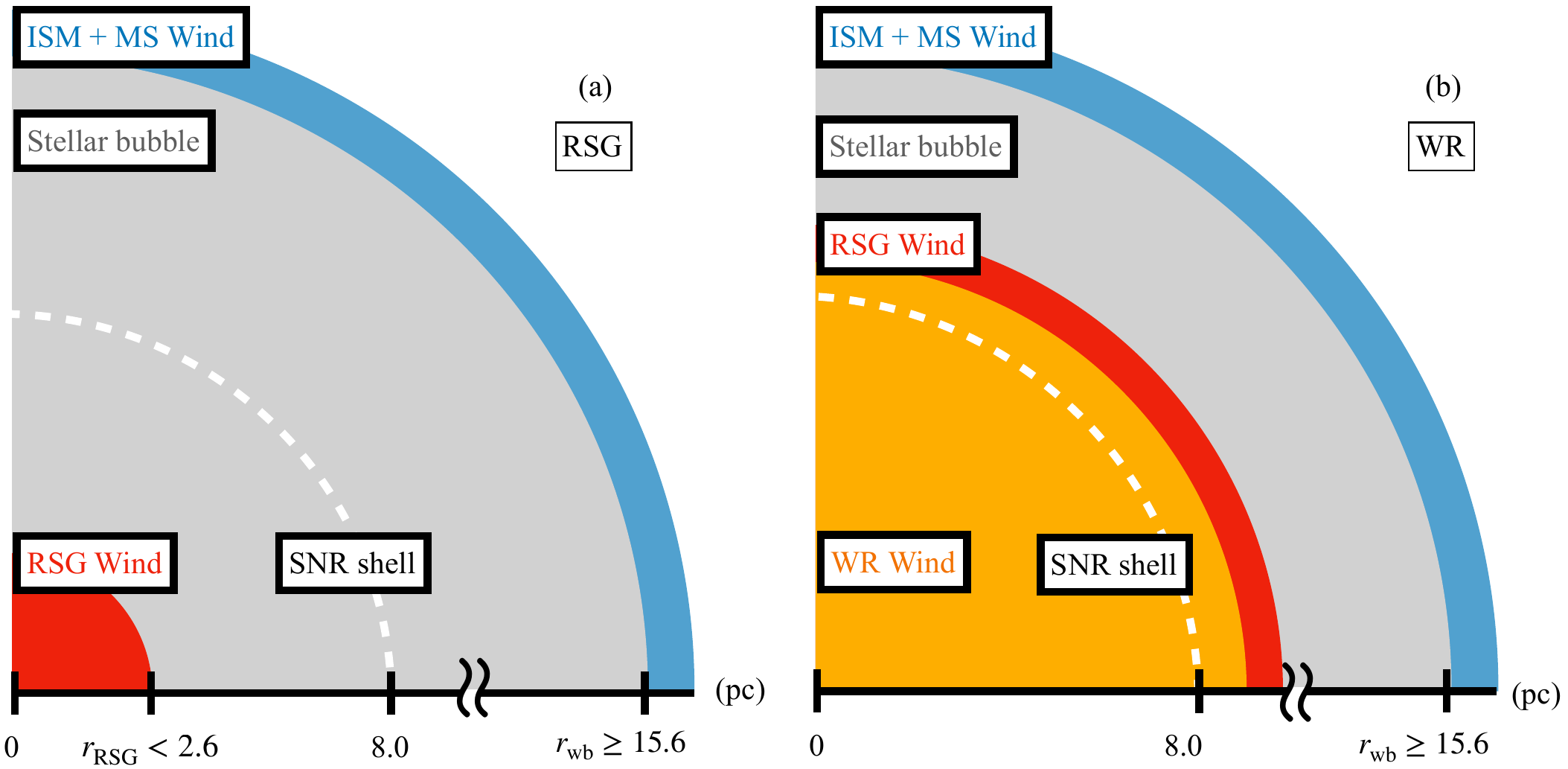}
 \end{center}
\caption{Schematic views of CSM before the SN explosion of the progenitor of \Gt. \textit{Left} (a): Case of a star ending its life as a RSG. Colors correspond to the shell of ISM and CSM produced through the MS phase (blue), stellar bubble (grey), and CSM produced through the RSG phase  (red). White dashed line represents the position of the forward shock of \Gt. \textit{Right} (b): Case of a star ending its life as a WR. Color corresponds are same as the left and the orange area represents the region dominated by CSM produced through WR phase.}
  \label{fig:wind}
\end{figure*}

\subsection{Progenitor Constraint with N/O}\label{subsec:N/O}
To constrain the progenitor parameters, we use a method established in our previous study \citep{Narita_2023}.
Figure~\ref{fig:NO} shows the relationship between the measured N/O and the progenitor parameters calculated with stellar evolution code of single stars \citep[Geneva code;][]{Ekstrom_2012} varying progenitor \M \ and initial rotation velocities ($v_{\rm{init}}/v_{\rm{crit}}=0, 0.4$, where $v_{\rm{init}}$ is the initial surface rotation velocity and $v_{\rm{crit}}$ is the critical velocity reaching when the gravitational acceleration is exactly counterbalanced by the centrifugal force).
We also vary the metallicity ($=1$ $Z_{\odot}$ and 0.5 $Z_{\odot}$) since \Gt \ is a young Galactic SNR, which is possibly in a solar- or slightly low-metallicity \citep[][]{Bhalerao_2019} environment.
Another parameter that can modify the value of N/O is a convective overshoot, which is the extrusion of convective motion due to non-zero velocities of material on a boundary between the convection zone and the radiative zone in stars.
The N-rich convection layer above the H-burning shell expands due to the strong convective overshoot and, therefore, the amount of N is larger than that of progenitor with the weak convective overshoot.
The N/O is enhanced in the case of the middle mass range (15--20~$M_{\odot}$) \citep[see Figure~8 of][]{Narita_2023}.
Another effect of the convective overshoot is an increase in the He core mass and, therefore, the minimum mass of a progenitor becoming an WR star is probably a few $M_{\odot}$ smaller than that of the progenitor with the weak overshoot \citep[cf][]{Yoshida_2019}.
\par

As shown in Figure~\ref{fig:NO}, we compared our result (N/O$=0.5\pm0.1$) with the above models.
The result indicates that  the progenitor of \Gt \ is likely a massive star with \M \ of $\gtrsim 30~M_{\odot}$ regardless of the rotation velocity and the metallicity.
It is inconsistent with most of the previous estimations of \M\ with ejecta abundances; 25--35~$M_{\odot}$ \citep{Park_2004, Kamitsukasa_2014}, $< 15~M_{\odot}$ \citep{Katsuda_2018}, 13--30~$M_{\odot}$ \citep{Bhalerao_2019}, and 12--16~$M_{\odot}$ \citep{Temim_2022}.
We thus investigate if the observed  low N/O could be explained by a relatively low-mass star experiencing strong mass-loss in a binary system, or something similar.
\par

\begin{figure*}[htbp]
 \begin{center}
  \includegraphics[width=180mm]{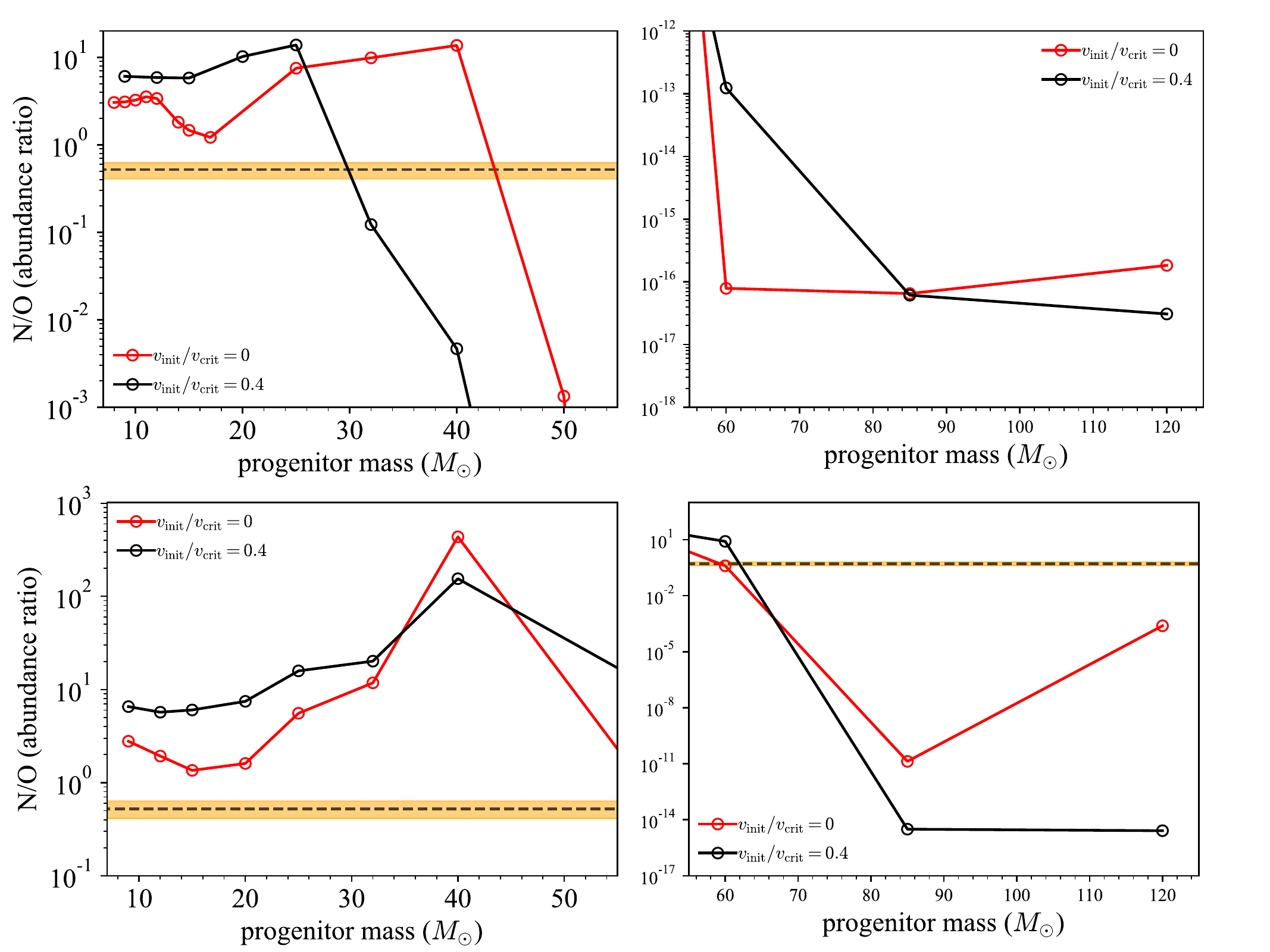}
 \end{center}
 \caption{\textit{Top}: N/O expected in the swept-up CSM with different \M \ with the solar metallicity. The black dashed lines  and the orange shaded region represent the averaged N/O in \Gt \ and its 1$\sigma$ error, respectively. The red and black circles correspond to cases of $v_{\rm{init}}/v_{\rm{crit}}=0$ and 0.4, respectively. \textit{Bottom}: The same as the top but in case of a lower metallicity (0.5 solar).}
\label{fig:NO}
\end{figure*}


Even in case of a lower \M \ progenitor  ($\lesssim 20~M_{\odot}$), if the mass-loss rate is somehow enhanced, a lager contribution of O-rich winds from He-burning products may be able to reduce the value of N/O.
One of the possibilities in this scenario is a gravitational effect of a companion star.
According to recent theories, if a progenitor is in a binary system, a star with \M$\lesssim 20~M_{\odot}$ can become a WR star with the mass of $\lesssim10~M_{\odot}$ \citep{Yoon_2010, Tauris_2015, Yoon_2017, Woosley_2019}.
It is noteworthy that such WR binary systems end as Type Ib/c SNe as these theories indicate.
This scenario resolves the discrepancy between our result based on the CSM composition ($\gtrsim 30~M_{\odot}$) and the other previous mass estimations based on the  ejecta composition ($\lesssim 30~M_{\odot}$).
It is also supported by a hydrodynamic simulation of \Gt, as presented by \citet{Temim_2022}, who argue that the progenitor of \Gt \ had a \M \  in the range of 12--16~$M_{\odot}$ and underwent a significant mass loss in a binary system.
From these results we conclude that the progenitor of \Gt \ was most likely a WR star in a binary system before the SN explosion rather than a high-\M \ single WR star.



\section{Conclusions}\label{sec:con}
We performed a high-resolution spectroscopy of \Gt \ with the RGS onboard XMM-Newton and robustly detected the \ion{N}{7} line for the first time.
Detailed spectral analysis indicates that the N-rich (${\rm N/Fe}\gtrsim1.3$), low-temperature component originates from shock-heated CSM produced by the CNO-cycle in the H-burning layer.
One of the intriguing findings is that \Gt \ has an O-rich CSM based on the measured abundance ratio of N/O ($=0.5\pm{0.1}$).
Comparing this  with the available stellar evolution codes, we found that none of the models for single low-mass progenitors can account for the observation; instead a high-mass star (\M$\gtrsim 30~M_{\odot}$; potentially a WR star) seems a more likely explanation.
Since the observed O-rich CSM implies a considerable mass loss, another possible scenario is a gravitational effect of a companion star in a binary system.
Many recent theories suggest that  a low-mass star (\M$\lesssim 20~M_{\odot}$) tends to end up as a WR star because of the mass stripping \citep{Yoon_2010, Yoon_2017, Woosley_2019, Tauris_2015}.
If this is the case, the mass-loss rate is significantly enhanced and the O-rich material is blown off, providing a natural explanation for the observed low N/O ratio and other previous studies on the progenitor mass estimation of \Gt.
Future studies of stellar and binary evolutions will allow us to constrain the progenitor parameters for remnants of such stripped-envelope SNe.
The method using the CNO element abundances  will be useful to reveal explosion mechanism of stripped SNe with upcoming microcalorimeter missions such as LEM \citep{Kraft_2022}, Athena \citep{Barret_2018}, and Lynx \citep{Gaskin_2019}.
\par
\section*{Acknowledgments}
We thank Mr. Ryohei Sato for helping the spectral analyses with RGS and Dr. Takaaki Tanaka and Mr. Shun Inoue for the meaningful discussion on the spectral analyses.
We also thank the anonymous referee for constructive comments.
This paper employs a list of Chandra datasets, obtained by the Chandra X-ray Observatory, contained in~\dataset[DOI:10.25574]{https://doi.org/10.25574/cdc.283}.
This work is supported by JSPS Core-to-Core Program grant No. JPJSCCA20220002 (T.N.) and JSPS/MEXT Science Research grant Nos. JP23KJ1350 (T.N.), JP19K03915, JP22H01265 (H.U.), JP21H04493 (T.G.T.), and JP20H00174, JP21H01121(SK).
\par

\bibliography{bibtex}{}
\bibliographystyle{aasjournal}

\end{document}